\documentstyle[preprint,aps]{revtex}
\begin{document}
\draft

\title{Non Linear Kinetics underlying Generalized Statistics}
\author{G. Kaniadakis}
\address{Dipartimento di Fisica - Politecnico di Torino -
Corso Duca degli Abruzzi 24, 10129 Torino, Italy \\
Istituto Nazionale di Fisica della Materia - Unit\'a del
 Politecnico di Torino }

\date{\today}
\maketitle

\begin {abstract} The purpose of the present effort is threefold.
Firstly, it is shown that there exists a principle, that we call
Kinetical Interaction Principle (KIP), underlying the non linear
kinetics in particle systems, independently on the picture
(Kramers, Boltzmann) used to describe their time evolution.
Secondly, the KIP imposes the form of the generalized entropy
associated to the system and permits to obtain the particle
statistical distribution, both as stationary solution of the non
linear evolution equation and as the state which maximizes the
generalized entropy. Thirdly, the KIP allows, on one hand, to
treat all the classical or quantum statistical distributions
already known in the literature  in a unifying scheme and, on the
other hand, suggests how we can introduce naturally new
distributions. Finally, as a working example of the approach to
the non linear kinetics here presented, a new non extensive
statistics is constructed and studied starting from a
one-parameter deformation of the exponential function holding the
relation $f(\!-x)f(x)=1$.
\end {abstract}


\section{Introduction}

In the last few decades there has been an intensive discussion on
non conventional classical or quantum statistics. Up to now
several entropies with the ensuing statistics have been
considered. For instance, in classical statistics, beside the
additive Boltzmann-Gibbs-Shannon entropy which leads to the
standard Maxwell-Boltzman statistics, one can find in the
literature, the entropies and/or related statistics introduced by
Druyvenstein \cite{D,D1}, Renyi \cite{R}, Sharma-Mittal \cite{SM},
Tsallis \cite{T}, Abe \cite{A}, Papa \cite{P}, Borges-Roditi
\cite{BR}, Landsberg-Vedral \cite{LV}, Anteneodo-Plastino
\cite{AP}, Frank-Daffertshofer \cite{FD}, among others. On the
other hand, in the literature we can find, besides the standard
Bose-Einstein and Fermi-Dirac quantum statistics and/or entropies,
the ones introduced by Gentile \cite{G}, Green \cite{GR},
Greenberg-Mohapatra \cite{GM}, Biedenharn \cite{BI}, Haldane-Wu
\cite{HW,HW1}, Acharya-Narayana Swamy \cite{AN},
Buyukkilic-Dimirhan \cite{BD} etc. This plethora of entropies
poses naturally some questions.

A first question is if it is possible and how to treat the above
entropies in the frame of a unifying context and from a more
general prospective, in such a way to distinguish the common
properties of the entropies, from the ones depending on the
particular form of the single entropy.

A second question is if it is possible to obtain the stationary
statistical distribution of the various non linear systems in the
frame of a time dependent scheme. Fermion and boson kinetics were
introduced in 1935 by Uehling and Uhlenbeck \cite{UU}. On the
other hand, the non linear kinetics associated with the anomalous
diffusion has been considered in the last few decades in several
papers by the mathematicians and by the physicists of the
condensed matter. After 1995, in the frame of the Fokker-Planck
picture, the anomalous diffusion has been linked with the time
dependent Tsallis statistical distribution
\cite{PP,TB,CJ1,CJ2,S,MPP,BO,BO1,BPPP,KL} and the kinetics of the
particles obeying the Haldane statistics \cite{NPB} and the quon
statistics \cite{PLA} has been considered.

The problem of the non linear kinetics from a more general point
of view has been considered only in 1994. In ref. \cite{KQBF} it
has been proposed an evolution equation (eq. (7) of the reference)
describing a generic non linear kinetics. Subsequently some
properties of this kinetics in the frame of the Fokker-Planck
picture has been studied  in ref. \cite{NPB} and in the frame of
the Boltzmann picture in ref. \cite{RK}. Finally, the kinetics
described by non linear Fokker-Planck equations has been
reconsidered recently in ref. \cite{FD,F}.

It is well known that the formalism used to describe the time
evolution of a statistical system, depends on the picture used to
describe the system. For instance, for a particle system
interacting with a bath, we can study its time evolution in the
phase space in the frame of the Kramers picture (Fokker-Planck
picture in the velocity space). Besides, for an isolated system,
we can study its time evolution in the phase space adopting the
Boltzmann picture.

A third question which arises at this point is if the entropy of a
system, or its stationary statistical distribution, depends and
how on the particular picture used to describe the system.

A fourth and last question is if it exists a principle underlying
the time evolution of the system, in the two pictures. Obviously
if this principle exists, it must define both the entropy and the
stationary statistical distribution of the system.

The present paper is concerned with the above questions. Its
principal goal is to show, that exists a principle in the
following called Kinetical Interaction Principle (KIP), which
governs the particle kinetics and imposes the form of the entropy
of the system independently on the particular picture used to
describe the system. Within the two pictures and in a unifying
context, the H-theorem is proved and the form of the the
generalized entropy together with the stationary statistical
distribution for a generic non linear system are obtained.

The paper is organized as it follows. In Sect. II,  we introduce
the KIP underlying the kinetics of a particle system, without
regard  if it interacts with its environment or if it is an
isolated system . In Sect. III, we study the nonlinear kinetics in
the Kramers picture implied by KIP. In particular, after writing
the evolution equation  of the system we obtain its entropy and
also its statistical distribution both as the stationary state of
the evolution equation and by using the maximum entropy principle.
The stability of this equilibrium distribution is also studied. In
Sect. IV, we examine an isolated particle system and describe its
nonlinear kinetics governed by the KIP in the Boltzmann picture.
In particular the equilibrium and the stability of the system are
studied. In Sect. V, we consider in an unifying context some
examples of already known classical and quantum statistical
distributions, in order to test and highlight  the utility of the
formalism here developed. In Sects. VI and VII, we consider, just
as a working example, a new statistical distribution, the
$\kappa$-deformed distribution, which arises naturally in the
frame of this formalism. This distribution is obtained according
to the KIP both as stationary solution of a nonlinear evolution
equation and by using the maximum entropy principle.  In Sect.
VIII, we consider some concrete physical systems where the
$\kappa$-deformed distribution can be adopted. Finally in Sect.
IX, some concluding remarks are reported.

\section{A principle underlying  the kinetics}

{\it Isolated systems:} Let us consider an isolated  system
composed by $N$ identical particles. We make the hypothesis that
the system is a low density gas so that we can describe it by a
one particle distribution function. The interaction of a particle
in the site $\mbox{\boldmath $r$}=(\mbox{\boldmath
$x$},\mbox{\boldmath $v$})$ where the particle density is
$f=f(t,\mbox{\boldmath $r$})$, with a second particle in the site
$\mbox{\boldmath $r$}_1=(\mbox{\boldmath $x$}_1,\mbox{\boldmath
$v$}_1)$ where the particle density is $f_1=f(t,\mbox{\boldmath
$r$}_1)$, changes the states of both the particles. After the
interaction we find the first particle to the site
$\mbox{\boldmath $r$}'=(\mbox{\boldmath $x$}',\mbox{\boldmath
$v$}')$, where the particle density is $f'=f(t,\mbox{\boldmath
$r$}')$ and the second particle to the site $\mbox{\boldmath
$r$}'_1=(\mbox{\boldmath $x$}'_1,\mbox{\boldmath $v$}'_1)$, where
the particle density is $f'_1=f(t,\mbox{\boldmath $r$}'_1)$. We
postulate that the transition probability from the state where the
particles occupy the sites $\mbox{\boldmath $r$}$ and
$\mbox{\boldmath $r$}_1$, after interaction, to the state where
the particles occupy the sites $\mbox{\boldmath $r$}'$ and
$\mbox{\boldmath $r$}'_1$  is given by
\begin{equation}
\!\pi(t,\mbox{\boldmath $r$}\!\!\rightarrow\!\!\mbox{\boldmath
$r$}'\!, \mbox{\boldmath $r$}_1\!\!\rightarrow\!\!\mbox{\boldmath
$r$}'_1)\!=\! T(t,\mbox{\boldmath $r$},\mbox{\boldmath
$r$}'\!,\mbox{\boldmath $r$}_1,\mbox{\boldmath $r$}'_1)\,
\gamma(f,\!f')\gamma(f_1,\!f'_1)\, . \label{N2}
\end{equation}
The first factor in (\ref{N2}) is the transition rate which
depends only on the nature of the two body particle interaction.
Then this factor is proportional to the cross section of the two
body interaction and doesn't depend on the particle population of
the four sites. Being the system composed by identical particles,
the second and third factors are given by the same function and
differ only on the arguments. The factor $\gamma(f,f')$ in
(\ref{N2}) is an arbitrary function of the particle populations of
the starting and of arrival sites. Eq.(\ref{N2}) takes into
account two body interactions and for this reason the function
$\gamma(f,f')$ must satisfy the condition $\gamma(0,f')=0 $
because, if the starting site is empty, the transition probability
is equal to zero. The dependence of the function $\gamma(f,f')$ on
the particle population $f'$ of the arrival site plays a very
important role in the particle kinetics because can stimulate or
inhibite the particle transition $\mbox{\boldmath
$r$}\rightarrow\mbox{\boldmath $r$}'$ in such a way that
interactions originated from collective effects can be taken into
account. The condition $\gamma(f,0) \neq 0$ requires that in the
case the arrival site is empty the transition probability must
depend only on the population of the starting site. We note that
for the standard linear kinetics the relation $\gamma(f,f')=f$
holds.

{\it Systems interacting with a bath:} We consider now the case of
a particle system interacting with its environment which we
consider as a bath. The particle which transits from the site
$\mbox{\boldmath $r$}$ to the site $\mbox{\boldmath $r$}'$ is now
the test particle while the one that transits from the site
$\mbox{\boldmath $r$}_1$ to the site $\mbox{\boldmath $r$}'_1$ is
a particle of the bath and the factor entering into the transition
probability is indicated with $\tilde \gamma \,(f_1,\!f'_1)$ and
depends on the nature of the bath. If we address our attention to
the test particle and after posing
\begin{equation}
\! W(t,\mbox{\boldmath $r$},\mbox{\boldmath $r$}')\!=\!\int d^{2n}
r_1 \, d^{2n} r'_1 \, T(t,\mbox{\boldmath $r$},\mbox{\boldmath
$r$}'\!,\mbox{\boldmath $r$}_1,\mbox{\boldmath $r$}'_1)\, \tilde
\gamma \,(f_1,\!f'_1),
\end{equation}
the transition probability (\ref{N2}) transforms immediately into
\begin{equation}
\pi(t,\mbox{\boldmath $r$}\rightarrow\mbox{\boldmath $r$}')=
W(t,\mbox{\boldmath $r$},\mbox{\boldmath $r$}') \, \gamma(f,f') \
\ . \label{N1}
\end{equation}
We remark that the transition rate $W(t,\mbox{\boldmath
$r$},\mbox{\boldmath $r$}')$ depends only on the nature of the
interaction between the test particle and the bath and doesn't
depend on the population of the test particle in the starting and
arrival sites.

{\it Kinetical interaction principle}: In this paper, we will
study kinetics coming from the transition probabilities (\ref{N1})
and (\ref{N2}) when the function $\gamma$ satisfies the condition
\begin{equation}
\frac{\gamma( f,f')}{\gamma( f',f)}=\frac{\kappa(f)}{\kappa(f')} \
\ , \label{N3}
\end{equation}
where $\kappa(f)$ is a positive real function. This condition
implies that $\gamma( f,f')/\kappa(f)$ is a symmetric function.
Then we can pose $\gamma( f,f')=\kappa(f)b(f)b(f')c(f,f')$ where
$b(f)$ and $c( f,f')=c(f',f)$ are two real arbitrary functions. It
will be convenient later on to introduce the real arbitrary
function $a(f)$ by means of
\begin{equation}
\kappa(f)=\frac{a(f)}{b(f)} \ \ , \label{N4}
\end{equation}
and write $\gamma( f,f')$ under the guise
\begin{equation}
\gamma(f,f')= a(f) \, b(f')\, c(f,f')\ \ . \label{N5}
\end{equation}
We claim at this point that $\gamma(f,f')$ given by (\ref{N5})
with $a(f)$ and $b(f)$ linked through (\ref{N4}), is the most
general function obeying the condition (\ref{N3}). We wish to note
that $\gamma(f,f')$ is given as a product of three factors. The
first factor  $a(f)$ is an arbitrary function of the particle
population of the starting site and satisfies the condition
$a(0)=0$ because if the starting site is empty the transition
probability is equal to zero. The second factor $b(f')$ is an
arbitrary function of the arrival site particle population. For
this function we have the condition $b(0)=1$ which requires that
the transition probability does not depend on the arrival site if,
in it, particles are absent. The expression of the function
$b(f')$ plays a very important role in the particle kinetics,
because stimulates or inhibites the transition $\mbox{\boldmath
$r$}\rightarrow\mbox{\boldmath $r$}'$, allowing in such a way to
consider interactions originated from collective effects. Finally,
the third factor $c(f,f')$ takes into account that the populations
of the two sites, namely $f$ and $f'$, can eventually affect the
transition, collectively and symmetrically.

The function $\gamma(f,f')$ given by (\ref{N5}) defines a special
interaction which involves, separately and/or together, the two
particle bunches entertained in the starting and arrival sites. We
observe that this interaction is different from the one depending
on the coordinates of the sites involved in the transition which
one takes into account by means of the functions $T$ (cross
section) in (\ref{N2}) or by $W$ (transition rate) in (\ref{N1}).
In order to explain the nature of the interaction introduced by
the function $\gamma(f,f')$ we start by considering the case
\begin{equation}
\gamma(f,f')= f (1- f') \ \ . \label{N55}
\end{equation}
It is well known \cite{UU,KQF} that this particular expression for
the $\gamma(f,f')$ given by (\ref{N55}) takes into account the
Pauli exclusion principle and defines completely the fermion
kinetics. Other expressions of the function $\gamma(f,f')$ take
into account interactions introduced by the generalized
exclusion-inclusion principle \cite{KQBF}, the Haldane generalized
exclusion principle \cite{HW,NPB}, the Tsallis principle
underlying the nonextensive statistics \cite{T,PP} etc. We observe
that  the above mentioned principles impose the form of the
collisional integral in the kinetic equations through the choice
of $\gamma(f,f')$. It is worth noting that in the cases of Haldane
statistics the particular expression of $\gamma(f,f')$ is
originated from the fractal structure of the single particle
Hilbert space, its dimension depending  on the particle number in
the considered state \cite{HW,NPB}. Also the Tsallis statistics is
originated from the fractal structure of the relevant particle
phase space \cite{T}.

Taking into account that particular choices of $\gamma(f,f')$
reproduce the  already known principles above mentioned, we can
see the function $\gamma(f,f')$ as describing a general principle
which we call {\it Kinetical Interaction Principle} (KIP). The KIP
defines a special collective interaction which could be very
useful to describe the dynamics of many body systems. As we will
see in the following sections, the KIP both governs the system
evolving toward the equilibrium and imposes the stationary state
of the system.

\section{Kramers generalized kinetics}

In the following we study the particle kinetics in a
$2n$-dimensional phase space of a dilute system composed by $N$
identical particles interacting with an equilibrated bath. The
procedure which we use in the present section to derive the
evolution equation of the system, is a generalization to the non
linear case of the standard procedure, involving the Kramers-Moyal
expansion and the first neighbor approximation, which was
introduced firstly to study the linear kinetics. We indicate with
$\mbox{\boldmath $x$}$ and $\mbox{\boldmath $v$}$ the position and
the velocity variables, respectively. The particles evolve under
an external potential $V=V(\mbox{\boldmath $x$})$. The evolution
equation for the distribution function $f=f(t,\mbox{\boldmath
$x$},\mbox{\boldmath $v$})$ is given by
\begin{equation}
\frac{d f}{d t}= \int_{\cal R} \left [ \pi(t,\mbox{\boldmath
$x$},\mbox{\boldmath $v$}'\rightarrow\mbox{\boldmath $v$})-
\pi(t,\mbox{\boldmath $x$},\mbox{\boldmath
$v$}\rightarrow\mbox{\boldmath $v$}') \right ] \, d^n v' \ \ ,
\label{N6}
\end{equation}
where $df/dt$ is the total time derivative while the transition
probability according to the KIP is given by
\begin{equation}
\pi(t,\mbox{\boldmath $x$},\mbox{\boldmath
$v$}\rightarrow\mbox{\boldmath $v$}')= W(t,\mbox{\boldmath
$x$},\mbox{\boldmath $v$},\mbox{\boldmath $v$}') \, \gamma(f,f') \
.
\end{equation}
Let us write the transition rate as $W(t,\mbox{\boldmath
$x$},\mbox{\boldmath $v$},\mbox{\boldmath
$v$}')=w(t,\mbox{\boldmath $x$},\mbox{\boldmath
$v$},\mbox{\boldmath $v$}'-\mbox{\boldmath $v$})$, where the last
argument in $w$ represents the change of the velocity during the
transition. In the following, for simplicity, we indicate
explicitly only the dependence on the velocity variables of the
functions $w(t,\mbox{\boldmath $x$},\mbox{\boldmath
$v$},\mbox{\boldmath $v$}'-\mbox{\boldmath $v$})$ and
$f(t,\mbox{\boldmath $x$},\mbox{\boldmath $v$})$. We start by
writing Eq.(\ref{N6}) as follows
\begin{eqnarray}
\frac{d f }{d t}&& = \int_{\cal R} w(\mbox{\boldmath
$v$}+\mbox{\boldmath $y$},\mbox{\boldmath $y$}) \,
\gamma[f(\mbox{\boldmath $v$}+\mbox{\boldmath
$y$}),f(\mbox{\boldmath $v$})] \, d^n y \nonumber
\\ && -\int_{\cal R} w(\mbox{\boldmath $v$},\mbox{\boldmath
$y$}) \, \gamma[f(\mbox{\boldmath $v$}),f(\mbox{\boldmath
$v$}-\mbox{\boldmath $y$})] \, d^n y \ \ . \label{A1}
\end{eqnarray}

For physical systems evolving very slowly $w(\mbox{\boldmath
$v$},\mbox{\boldmath $y$})$ decreases very expeditiously as
$\mbox{\boldmath $y$}$ increases and we can consider only the
transitions for which $\mbox{\boldmath $v$}\pm\mbox{\boldmath
$y$}\approx \mbox{\boldmath $v$}$. At this point we make use of
the two following Taylor expansions
\begin{eqnarray}
&&w(\mbox{\boldmath $v$}+\mbox{\boldmath $y$},\mbox{\boldmath
$y$}) \, \gamma[f(\mbox{\boldmath $v$}+\mbox{\boldmath
$y$}),f(\mbox{\boldmath $v$})]\nonumber \\
&&=\!\sum_{m=0}^{\infty}\!\frac{1}{m!}\left[\frac{\partial^m
\{w(\mbox{\boldmath $u$},\mbox{\boldmath $y$}) \,
\gamma[f(\mbox{\boldmath $u$}),f(\mbox{\boldmath
$v$})]\}}{\partial u_{\alpha_1}\partial u_{\alpha_2}...\partial
u_{\alpha_m}}\right]_{_{\mbox{\boldmath $u$}=\mbox{\boldmath
$v$}}} \!\!y_{\alpha_1}y_{\alpha_2}...y_{\alpha_m}, \nonumber
\end{eqnarray}
\begin{eqnarray}
&& \gamma[f(\mbox{\boldmath $v$}),f(\mbox{\boldmath
$v$}-\mbox{\boldmath $y$})]\nonumber \\
&&=\sum_{m=0}^{\infty}\frac{(-1)^m}{m!}\left[\frac{\partial ^m
\gamma[f(\mbox{\boldmath $v$}),f(\mbox{\boldmath $u$})]}{\partial
u_{\alpha_1}\partial u_{\alpha_2}...\partial
u_{\alpha_m}}\right]_{_{\mbox{\boldmath $u$}=\mbox{\boldmath
$v$}}}y_{\alpha_1}y_{\alpha_2}...y_{\alpha_m} , \nonumber
\end{eqnarray}
and after substitution in (\ref{A1}) we obtain the following
Kramers-Moyal expansion.
\begin{eqnarray}
&&\frac{d f(t,\mbox{\boldmath $x$},\mbox{\boldmath $v$}) }{d t}
=\sum_{m=1}^{\infty}\Bigg[\frac{\partial^m
\{\zeta_{\alpha_1\alpha_2...\alpha_m}(t,\mbox{\boldmath
$x$},\mbox{\boldmath $u$}) \, \gamma[f(t,\mbox{\boldmath
$x$},\mbox{\boldmath $u$}),f(t,\mbox{\boldmath
$x$},\mbox{\boldmath $v$})]\}}{\partial u_{\alpha_1}\partial
u_{\alpha_2}...\partial u_{\alpha_m}} \nonumber \\ &&+ (-1)^{m-1}
\zeta_{\alpha_1\alpha_2...\alpha_m}(t,\mbox{\boldmath
$x$},\mbox{\boldmath $v$})\frac{\partial^m
\gamma[f(t,\mbox{\boldmath $x$},\mbox{\boldmath
$v$}),f(t,\mbox{\boldmath $x$},\mbox{\boldmath $u$})]}{\partial
u_{\alpha_1}\partial u_{\alpha_2}...\partial
u_{\alpha_m}}\Bigg]_{_{\mbox{\boldmath $u$}=\mbox{\boldmath $v$}}}
,\label{N601}
\end{eqnarray}
where the $m$-th order momentum
$\zeta_{\alpha_1\alpha_2...\alpha_m}(t,\mbox{\boldmath
$x$},\mbox{\boldmath $v$})$ of the transition rate is defined as:
\begin{equation}
\zeta_{\alpha_1\alpha_2...\alpha_m}(t,\mbox{\boldmath
$x$},\mbox{\boldmath $v$})=\frac{1}{m!}\int_{\cal
R}y_{\alpha_1}y_{\alpha_2}...y_{\alpha_m} w(t,\mbox{\boldmath
$x$},\mbox{\boldmath $v$},\mbox{\boldmath $y$}) d^n y .
\end{equation}
We remark that from Eq. (\ref{N601}), where the dependence on all
the variables $t,\mbox{\boldmath $x$},\mbox{\boldmath $v$}$ is
indicated explicitly, we can obtain as a particular case Eq. (7)
of ref. \cite{KQBF}.

In the frame of the first neighbor approximation only the first
order (drift coefficient) $\zeta_{i}$ and the second order
(diffusion coefficient) $\zeta_{ij}$ momenta of the transition
rate are considered. Indicating again explicitly only the
dependence on the velocity variables Eq.(\ref{N601}) reduces to
the following non linear second order partial differential
equation
\begin{eqnarray}
&&\frac{d f}{d t} =\Bigg[ \frac{\partial\{\zeta_i(\mbox{\boldmath
$u$})\gamma[f(\mbox{\boldmath $u$}),f(\mbox{\boldmath
$v$})]\}}{\partial u_i} +\zeta_i(\mbox{\boldmath
$v$})\frac{\partial \gamma[f(\mbox{\boldmath
$v$}),f(\mbox{\boldmath $u$})] }{\partial u_i}  \nonumber
\\ &&+
\frac{\partial^2\{\zeta_{ij}(\mbox{\boldmath
$u$})\gamma[f(\mbox{\boldmath $u$}),f(\mbox{\boldmath
$v$})]\}}{\partial u_i \partial u_j}-\zeta_{ij}(\mbox{\boldmath
$v$}) \frac{\partial ^2 \gamma[f(\mbox{\boldmath
$v$}),f(\mbox{\boldmath $u$})]}{\partial u_i \partial u_j}
\Bigg]_{_{\mbox{\boldmath $u$}=\mbox{\boldmath $v$}}},
\end{eqnarray}
which after taking into account the two identities:
\begin{eqnarray}
&& \Bigg[ \frac{\partial\{\zeta_i(\mbox{\boldmath
$u$})\gamma[f(\mbox{\boldmath $u$}),f(\mbox{\boldmath
$v$})]\}}{\partial u_i}+\zeta_i(\mbox{\boldmath
$v$})\frac{\partial \gamma[f(\mbox{\boldmath
$v$}),f(\mbox{\boldmath $u$})] }{\partial
u_i}\Bigg]_{_{\mbox{\boldmath $u$}=\mbox{\boldmath $v$}}}
\nonumber \\ && = \frac{\partial}{\partial
v_i}\{\zeta_i(\mbox{\boldmath $v$})\gamma[f(\mbox{\boldmath
$v$}),f(\mbox{\boldmath $v$})]\}, \nonumber
\end{eqnarray}
and
\begin{eqnarray}
&& \Bigg[\frac{\partial^2\{\zeta_{ij}(\mbox{\boldmath
$u$})\gamma[f(\mbox{\boldmath $u$}),f(\mbox{\boldmath
$v$})]\}}{\partial u_i \partial u_j}-\zeta_{ij}(\mbox{\boldmath
$v$}) \frac{\partial ^2 \gamma[f(\mbox{\boldmath
$v$}),f(\mbox{\boldmath $u$})]}{\partial u_i \partial u_j}
\Bigg]_{_{\mbox{\boldmath $u$}=\mbox{\boldmath $v$}}} \nonumber \\
&& =\frac{\partial}{\partial v_i}\Bigg\{ \frac{\partial
\zeta_{ij}(\mbox{\boldmath $v$})}{\partial v_j}
\gamma[f(\mbox{\boldmath $v$}),f(\mbox{\boldmath $v$})]
 \nonumber \\ &&  + \zeta_{ij}(\mbox{\boldmath $v$})
\Bigg[\frac{\partial \gamma[f(\mbox{\boldmath
$u$}),f(\mbox{\boldmath $v$})]}{\partial u_j }-\frac{\partial
\gamma[f(\mbox{\boldmath $v$}),f(\mbox{\boldmath $u$})] }{\partial
u_j }\Bigg]_{_{\mbox{\boldmath $u$}=\mbox{\boldmath $v$}}}
\Bigg\}, \nonumber
\end{eqnarray}
assumes the form
\begin{eqnarray}
\frac{d f }{d t}=\frac{\partial}{\partial v_i}\Bigg\{ \Bigg
[\zeta_i(\mbox{\boldmath $v$})+\frac{\partial
\zeta_{ij}(\mbox{\boldmath $v$})}{\partial v_j} \Bigg ]
\gamma[f(\mbox{\boldmath $v$}),f(\mbox{\boldmath $v$})] \nonumber
\\ + \zeta_{ij}(\mbox{\boldmath $v$}) \Bigg[\frac{\partial
\gamma[f(\mbox{\boldmath $u$}),f(\mbox{\boldmath $v$})]}{\partial
u_j }-\frac{\partial \gamma[f(\mbox{\boldmath
$v$}),f(\mbox{\boldmath $u$})] }{\partial u_j
}\Bigg]_{_{\mbox{\boldmath $u$}=\mbox{\boldmath $v$}}} \Bigg\}.
\label{A4}
\end{eqnarray}
Finally Eq. (\ref{A4}) can be rewritten as
\begin{equation}
\frac{d f}{d t} = \frac{\partial}{\partial v_{i}} \! \left[
 \left (\zeta_{i}+\frac{\partial \zeta_{ij} }
{\partial v_{j}} \right )\gamma(f) +\zeta_{ij}\gamma(f)\lambda(f)
\frac{\partial f}{\partial v_{j}} \! \right ] \ \ , \label{N7}
\end{equation}
with $\gamma(f)= \gamma( f, \,f)$ and
\begin{eqnarray}
\lambda(f)=\left[\frac{\partial}{\partial f}\ln \frac{\gamma( f,
f')} {\gamma(f',f)}\right]_{f'=f}
 \ \ . \nonumber
\end{eqnarray}
By taking into account the condition (\ref{N3}), the function
$\lambda(f)$ simplifies as
\begin{equation}
\lambda(f)=\frac{\partial \ln \kappa(f)}{\partial f}  \ \ ,
\end{equation}
and Eq.(\ref{N7}) becomes
\begin{equation}
\!\! \frac{d f}{d t}\! =\! \frac{\partial}{\partial v_{i}} \!
\left[
 \!\left (\zeta_{i}+\frac{\partial \zeta_{ij} }
{\partial v_{j}} \right )\!\gamma(f)\! +\!\gamma(f) \frac{\partial
\ln \kappa(f)}{\partial f}\,\zeta_{ij} \frac{\partial f}{\partial
v_{j}} \! \right ] . \label{N8}
\end{equation}

We assume the independence of motion among the $n$ directions of
the homogeneous and isotropic $n$-dimensional velocity space and
pose $\zeta_{i}=J_i$, $\zeta_{ij}=D\delta_{ij}$, being
$\mbox{\boldmath $J$}$ and $D$ the drift and diffusion
coefficients, respectively. Moreover we introduce the function $U$
by means of
\begin{equation}
\beta \frac{\partial U}{\partial \mbox{\boldmath$v$}} =
\frac{1}{D} \left (\mbox{\boldmath$J$}+ \frac{\partial D }
{\partial \mbox{\boldmath$v$}} \right ) \ \ ,
\end{equation}
with $\beta$ a constant. In the following we will consider the
case where $U=U(\mbox{\boldmath $v$})$ depends exclusively on the
velocity. Taking into account that the potential
$V=V(\mbox{\boldmath $x$})$ depends only on the spatial variable,
Eq. (\ref{N8}) can be written as
\begin{equation}
 \frac{d f(t,\mbox{\boldmath$x$},\mbox{\boldmath$v$})}{d t}= \frac{\partial}{\partial \mbox{\boldmath$v$}}
\left (D(\mbox{\boldmath$v$}) \gamma(f)\frac{\partial}{\partial
\mbox{\boldmath$v$}} \bigg \{ \beta\left[
V(\mbox{\boldmath$x$})\!+\!U(\mbox{\boldmath$v$})\!-\!\mu
\right]\!+\! \ln \kappa(f) \bigg \} \! \right ), \label{N8a}
\end{equation}
with $\mu$ a constant and the expression of the total time
derivative given by
\begin{eqnarray}
\frac{d}{d t}=\frac{\partial}{\partial t}+
\frac{1}{m}\frac{\partial U(\mbox{\boldmath$v$})}{\partial
\mbox{\boldmath $v$}}\frac{\partial }{\partial \mbox{\boldmath
$x$}}-\frac{1}{m}\frac{\partial V(\mbox{\boldmath$x$})}{\partial
\mbox{\boldmath $x$}}\frac{\partial }{\partial \mbox{\boldmath
$v$}} \ \ . \nonumber
\end{eqnarray}
Equation (\ref{N8a}) represents the evolution equation of the
particle system in the Kramers picture and describe a non linear
kinetics. This non linear evolution equation can be written in the
form
\begin{eqnarray}
\frac{d f}{d t}+\frac{\partial}{\partial \mbox{\boldmath$v$}} \!
\left [D\gamma(f)\frac{\partial}{\partial \mbox{\boldmath$v$}}
\frac{\delta {\cal K}}{\delta f}\! \right ] = 0 \ \ , \label{N15}
\end{eqnarray}
where $\delta{\cal K}/\delta f$ is the functional derivatives of
the functional ${\cal K}$ defined through
\begin{equation}
{\cal K}= -\int_{\cal R}\ d^n x d^nv \,\, \int df  \ln
\frac{\kappa (f)}{\kappa (f_{\!s})} \ \ , \label{M3}
\end{equation}
where the stationary distribution $f_{\!s}=
f(\infty,\mbox{\boldmath$x$},\mbox{\boldmath$v$})$ is defined
through:
\begin{equation}
\ln \kappa (f_{\!s}) =-\beta \left[ V(\mbox{\boldmath
$x$})+U(\mbox{\boldmath$v$})-\mu\right] \ \ .
\end{equation}
We remark now that from Eq. (\ref{N15}) it follows immediately
that the stationary distribution maximizes ${\cal K}$ and can be
obtained from a variational principle:
\begin{eqnarray}
\frac{\delta {\cal K}}{\delta f}=0  \ \ \ \Rightarrow \ \ \
\frac{d f}{d t}= 0 \ \ ;
  \ \  f=f_{\!s}  \ \ . \nonumber
\end{eqnarray}

It is easy to verify that the functional ${\cal K}$ increases in
time
\begin{eqnarray}
\frac{d{\cal K}}{d t} \ &&= \int_{\cal R} d^n x d^nv \frac{\delta
{\cal K}}{\delta f} \frac{d f}{d t}\nonumber \\ \nonumber &&=
-\int_{\cal R}d^n x d^n v \frac{\delta {\cal K}}{\delta
f}\frac{\partial}{\partial \mbox{\boldmath$v$}} \!
\left[D\gamma(f)\frac{\partial} {\partial \mbox{\boldmath$v$}}
\frac{\delta {\cal K}}{\delta f} \right ] \\
 &&= \int_{\cal R}d^n x d^n v
D\gamma(f)\left(\frac{\partial}{\partial \mbox{\boldmath $v$}}
\frac{\delta {\cal K}}{\delta f} \right)^2  \geq 0 \ \ \label{H1}
.
\end{eqnarray}
In order to study the behaviour of the functional ${\cal K}(t)$
when $t\rightarrow\infty$ we introduce the function $\sigma(f)= -
\int d f \, \ln \kappa (f)$ so that $\kappa (f)$ can be written as
$\kappa (f)=\exp\left[-d \sigma/d f\right]$. Now we are able to
calculate, in the limit $t\rightarrow\infty$, the following
difference
\begin{eqnarray}
{\cal K}(t)\!-\!{\cal K}(\infty)=&& \int_{\cal R}d^n x d^nv \
[\sigma(f)-\sigma(f_{\!s})+(f-f_{\!s})\ln \kappa (f_{\!s})]
\nonumber \\ = && \int_{\cal R} d^n x
d^nv\left[\sigma(f)-\sigma(f_{\!s})-(f-f_{\!s}) \frac{d \sigma
(f_{\!s})}{d f_{\!s} }\right] \nonumber \\
 \approx && \int_{\cal R}d^n x d^nv\left[\frac{1}{2}\frac{d^2 \sigma
(f_{\!s})}{d f^2_{\!s} }(f-f_{\!s})^2 \right] \ \ , \label{H2}
\end{eqnarray}
and assume that $d^2 \sigma (f)/d f^2  \leq 0$. This requirement
is satisfied if  the function $\kappa(f)$ obeys to the condition
$d \kappa (f)/d f \geq 0$ and consequently we have ${\cal
K}(t)\leq {\cal K}(\infty)$, then ${\cal K}$ assumes its maximum
value for $t=\infty$. The inequalities $d{\cal K}(t)/dt\geq 0$ and
${\cal K}(t)\leq {\cal K}(\infty)$ imply that $-{\cal K}$ is a
Lyapunov functional and demonstrate the H-theorem. The functional
${\cal K}$ is the constrained entropy of the system and results to
be the sum of two terms: ${\cal K}=S + S_c$ where
\begin{equation}
S= - \int_{\cal R} d^n xd^nv \ \int d f \, \ln \kappa (f) \ \ ,
\label{M5}
\end{equation}
is the entropy of the system and $S_c\!=\!-\beta (E-\mu N)$. The
energy  $E$ of the system is given by
\begin{equation}
E= \int_{\cal
R}d^nxd^nv\left[V(\mbox{\boldmath$x$})+U(\mbox{\boldmath$v$})\right]
f \ \ .
\end{equation}
We remark that, being $\kappa(f)$ an arbitrary function, the
H-theorem has been verified in a unified way, for a very large
class of non linear systems interacting with a bath.

\section{Boltzmann generalized kinetics}

In the diffusive approximation, adopted in the previous section to
describe the changes of the particle states, the system is coupled
with its environment. In this frame, meanwhile the particle
system, interacting  with the bath, evolves toward the
equilibrium, both its entropy $S$ and ${\cal K}$ increases
monotonicaly. In the present section we will consider the particle
system isolated from its environment and describe its time
evolution in the frame of the more rigorous Boltzmann picture. In
presence of external forces derived from a potential the total
time derivative is defined as
\begin{eqnarray}
\frac{d}{d t}=\frac{\partial}{\partial t}+ \mbox{\boldmath
$v$}\frac{\partial }{\partial \mbox{\boldmath
$x$}}-\frac{1}{m}\frac{\partial V(\mbox{\boldmath $x$})}{\partial
\mbox{\boldmath $x$}}\frac{\partial }{\partial \mbox{\boldmath
$v$}} \ \ , \nonumber
\end{eqnarray}
and the evolution equation assumes the form:
\begin{eqnarray}
\frac{d f}{d t}= \int_{\cal R} d^nv'd^nv_1d^nv'_1  \big
[\pi(t,\mbox{\boldmath $x$},\mbox{\boldmath
$v$}'\rightarrow\mbox{\boldmath $v$}, \mbox{\boldmath
$v$}'_1\rightarrow\mbox{\boldmath $v$}_1)&& \nonumber \\
-\pi(t,\mbox{\boldmath $x$},\mbox{\boldmath
$v$}\rightarrow\mbox{\boldmath $v$}', \mbox{\boldmath
$v$}_1\rightarrow\mbox{\boldmath $v$}'_1)&& \big ] \ \ .
\label{N16}
\end{eqnarray}
Eq. (\ref{N16}) describes a non linear generalized kinetics,  the
transition probabilities being defined, according to the KIP, as:
\begin{eqnarray}
&&\pi(t,\mbox{\boldmath $x$},\mbox{\boldmath
$v$}\rightarrow\mbox{\boldmath $v$}', \mbox{\boldmath
$v$}_1\rightarrow\mbox{\boldmath $v$}'_1)\nonumber \\ && =
T(t,\mbox{\boldmath $x$},\mbox{\boldmath $v$},\mbox{\boldmath
$v$}',\mbox{\boldmath $v$}_1,\mbox{\boldmath $v$}'_1)\,
\gamma(f,f')\gamma(f_1,f'_1) \ \ . \label{N17}
\end{eqnarray}
In (\ref{N17}) we have posed \hbox{$f=f(t,\mbox{\boldmath
$x$},\mbox{\boldmath $v$})$}, \hbox{$f'=f(t,\mbox{\boldmath
$x$},\mbox{\boldmath $v$}')$} and analogously
\hbox{$f_1=f(t,\mbox{\boldmath $x$},\mbox{\boldmath $v$}_1)$},
\hbox{$f'_1=f(t,\mbox{\boldmath $x$},\mbox{\boldmath $v$}'_1)$} in
order to consider point-like binary collisions. A symmetry is
imposed to \hbox{$T=T(t,\mbox{\boldmath $x$},\mbox{\boldmath
$v$},\mbox{\boldmath $v$}',\mbox{\boldmath $v$}_1,\mbox{\boldmath
$v$}'_1)$} by the principle of detailed balance $
T(t,\mbox{\boldmath $x$},\mbox{\boldmath $v$},\mbox{\boldmath
$v$}',\mbox{\boldmath $v$}_1,\mbox{\boldmath
$v$}'_1)=T(t,\mbox{\boldmath $x$},\mbox{\boldmath
$v$}',\mbox{\boldmath $v$},\mbox{\boldmath $v$}'_1,\mbox{\boldmath
$v$}_1)$, so that Eq. (\ref{N16}) assumes the form
\begin{eqnarray}
\frac{d f}{d t}  = \int_{\cal R} d^nv'd^nv_1d^nv'_1 \
T(t,\mbox{\boldmath $x$},\mbox{\boldmath $v$},\mbox{\boldmath
$v$}',\mbox{\boldmath $v$}_1,\mbox{\boldmath $v$}'_1)&&  \nonumber
\\ \ \times \big [ \gamma(f',f)\gamma(f'_1,f_1) -
\gamma(f,f')\gamma(f_1,f'_1) \big ]&& \ \ . \label{N18}
\end{eqnarray}
Alternatively, by taking into account (\ref{N5}), one can write:
\begin{eqnarray}
\frac{d f}{d t} && = \! \! \int_{\cal R}\! \! d^nv'd^nv_1d^nv'_1 \
T \ c(f,f')c(f_1,f'_1) \nonumber
\\ && \times \big [a(f')b(f)a(f'_1)b(f_1) -
a(f)b(f')a(f_1)b(f'_1) \big ] \ \ . \label{N1801}
\end{eqnarray}
We note that when \hbox{$c(f,f')=c(f_1,f'_1)=1$}, Eq.
(\ref{N1801}) reduces to the equation recently considered in ref.
\cite{RK}. In the following we will describe the system using Eq.
(\ref{N18}) which can be rewritten as
\begin{equation}
\frac{d f}{d t}= \int_{\cal R} d^nv'd^nv_1d^nv'_1 \ T \ {\cal Q}\
\gamma(f',f)\ \gamma(f'_1,f_1) \ \ , \label{N19}
\end{equation}
where  the auxiliary function ${\cal Q}={\cal
Q}(f,f',f_1,f'_1)\leq 1$ is defined as
\begin{eqnarray}
{\cal Q}=1-\frac{\gamma(f,f')\gamma(f_1,f'_1)}{
\gamma(f',f)\gamma(f'_1,f_1)} \ \ . \nonumber
\end{eqnarray}
Taking into account the condition (\ref{N3}), the function ${\cal
Q}$ becomes ${\cal Q}=1-[\kappa(f)\kappa(f_1)]/
[\kappa(f')\kappa(f'_1)]$ and can be written immediately in the
following form
\begin{eqnarray}
{\cal Q}=
1-\exp\big[\ln\kappa(f)+\ln\kappa(f_1)-\ln\kappa(f')-\ln\kappa(f'_1)
\big] \nonumber \ .
\end{eqnarray}
We consider now the system at the equilibrium. From the evolution
equation (\ref{N19}) we have $df/dt=0$, and ${\cal Q}=0$. The
first condition $df/dt=0$ taking into account the definition of
the total time derivative, implies for the stationary distribution
$f=f_{\!s}$ that $f_{\!s}= f_{\!s}[m\mbox{\boldmath$v$}^2/2 +
V(\mbox{\boldmath$x$})]$. On the other hand from the condition
${\cal Q}=0$ we have
\begin{eqnarray}
\ln\kappa(f_{\!s})+\ln\kappa(f_{\!s1})-
\ln\kappa(f'_{\!s})-\ln\kappa(f'_{\!s1})=0 \ . \nonumber
\end{eqnarray}
This last equation allows us to conclude that the quantity
$\ln\kappa(f_{\!s})$ is a collisional invariant for the particle
system. If we suppose that the binary interparticle collisions
conserve the particle number and the kinetic energy
\begin{eqnarray}
\frac{1}{2}m\mbox{\boldmath$v$}^2 +
\frac{1}{2}m\mbox{\boldmath$v$}_1^2 =
\frac{1}{2}m{\mbox{\boldmath$v$}'}^2 +
\frac{1}{2}m{\mbox{\boldmath$v$}'_1}^2 \ \ , \nonumber
\end{eqnarray}
we have that the quantity
$-\beta\left[\frac{1}{2}m\mbox{\boldmath$v$}^2+
V(\mbox{\boldmath$x$})-\mu\right]$ is the more general collisional
invariant.  Then we obtain the condition
\begin{equation}
\ln \kappa(f_{\!s})=-\beta\left[\frac{1}{2}m\mbox{\boldmath$v$}^2+
V(\mbox{\boldmath$x$})-\mu\right] \ \ , \label{M4}
\end{equation}
which defines the stationary distribution, so that ${\cal Q}$
becomes
\begin{eqnarray}
{\cal Q}\!=\! 1\!-\!\exp \!
\left[\ln\frac{\kappa(f)}{\kappa(f_{\!s})}\!+\!
\ln\frac{\kappa(f_1)}{\kappa(f_{\!s1})}
\!-\!\ln\frac{\kappa(f')}{\kappa(f'_{\!s})}\!-\!
\ln\frac{\kappa(f'_1)}{\kappa(f'_{\!s1})} \right] \ . \nonumber
\end{eqnarray}
After  introducing the functional ${\cal K}$ by means of
(\ref{M3}) with $f_{\!s}$ is given by (\ref{M4}) the quantity
${\cal Q}$ can be written in the form:
\begin{eqnarray}
{\cal Q}= 1-\exp\left( - \frac{\delta{\cal K}[f]}{\delta f} -
\frac{\delta{\cal K}[f_1]}{\delta f_1} + \frac{\delta{\cal
K}[f']}{\delta f'}  + \frac{\delta{\cal K}[f'_1]}{\delta f'_1}
 \right) \ . \nonumber
\end{eqnarray}
Finally the evolution equation (\ref{N19}) assumes the form
\begin{eqnarray}
&&\frac{d f}{d t} = \int_{\cal R} d^nv'd^nv_1d^nv'_1 \
 T \ \ \gamma(f',f)\ \gamma(f'_1,f_1) \nonumber \\
 && \times \left \{\! 1\!-\!\exp\left( \!-
\frac{\delta{\cal K}[f]}{\delta f}\! -\!\frac{\delta{\cal
K}[f_1]}{\delta f_1}\! + \!\frac{\delta{\cal K}[f']}{\delta f'}\!+
\!\frac{\delta{\cal K}[f'_1]}{\delta f'_1} \right )\! \right \} \
. \label{N20}
\end{eqnarray}
From the structure of (\ref{N20}) we have that the stationary
distribution $f_{\!s}=
f(\infty,\mbox{\boldmath$x$},\mbox{\boldmath$v$})$ can be obtained
from a variational principle
\begin{eqnarray}
\frac{\delta {\cal K}}{\delta f}=0 \ \ \ \Rightarrow \ \ \ \frac{d
f}{d t}=0 \ \ ; \ \ f=f_{\!s} \ \ . \nonumber
\end{eqnarray}
We study now how  the functional ${\cal K}$ evolves in time. Its
time derivative is given by
\begin{eqnarray}
\frac{d{\cal K}}{d t}&&= - \int_{\cal R} d^nxd^nv \ \ln
\frac{\kappa(f)}{\kappa(f_{\!s})} \frac{d f}{d t} \nonumber \\
&&=- \int_{\cal R} d^nxd^nvd^nv'd^nv_1d^nv'_1  \ T \ \ln
\frac{\kappa(f)}{\kappa(f_{\!s})} \nonumber
\\ &&\times  [\gamma(f',f)\
\gamma(f'_1,f_1)-\gamma(f,f')\ \gamma(f_1,f'_1)] \ . \label{N21}
\end{eqnarray}
The symmetry of the integrand function permit us to write
(\ref{N21}) as it follows
\begin{eqnarray}
\frac{d{\cal K}}{d t}&&= -\int_{\cal R} d^nxd^nvd^nv'd^nv_1d^nv'_1
\ \frac{1}{4}\ T \ \nonumber
\\ &&\times \left[ \ln\frac{\kappa(f)}{\kappa(f_{\!s})}+
\ln\frac{\kappa(f_1)}{\kappa(f_{\!s1})}
-\ln\frac{\kappa(f')}{\kappa(f'_{\!s})}-
\ln\frac{\kappa(f'_1)}{\kappa(f'_{\!s1})} \right] \nonumber \\
&&\times \ [\gamma(f',f)\ \gamma(f'_1,f_1)-\gamma(f,f')\
\gamma(f_1,f'_1)] \ \ .
\end{eqnarray}
After taking into account the expressions of ${\cal Q}$ we have:
\begin{eqnarray}
\frac{d{\cal K}}{d t}&&= \int_{\cal R} d^nxd^nvd^nv'd^nv_1d^nv'_1
\ \frac{1}{4}\ T
 \nonumber
\\ &&\times  \ [-{\cal Q}\ln (1-{\cal
Q})]\ \gamma(f',f)\ \gamma(f'_1,f_1) \ . \label{N22}
\end{eqnarray}
We observe now that ${\cal Q}\leq 1$ and then we have $ -{\cal
Q}\ln (1-{\cal Q})\geq 0$. This implies that the integrand
function in (\ref{N22}) is a non negative function. Then we
conclude that \hbox{$d {\cal K}/d t \geq 0$}. Starting from the
definition of ${\cal K}$ and following the procedure adopted in
Sect. III we obtain ${\cal K}(t) \leq {\cal K}(\infty)$. We can
conclude at this point that $-{\cal K}$ is a Lyapunov functional.
It is easy to verify that
\begin{equation}
{\cal K}=S - \beta (E-\mu N) \ \ , \label{N23}
\end{equation}
where the entropy of the system $S$ is defined through (\ref{M5})
while its energy $E$ is given by:
\begin{equation}
E= \int_{\cal R}d^nxd^nv\left[\frac{1}{2}m
\mbox{\boldmath$v$}^2+V(\mbox{\boldmath$x$})\right] f \ \ ,
\end{equation}
which is a conserved quantity,  $d E/d t= 0$, as the particle
number $N$. Consequently (\ref{N23}) can be written as
\begin{equation}
{\cal K}(t)=S(t) + {\rm constant} .
\end{equation}
The H-theorem for the isolated nonlinear system follows
immediately:
\begin{equation}
\frac{d S}{d t} \geq 0 \ \ \ \ ; \ \ \ \ S(t) \leq S(\infty) \ \ .
\end{equation}

\section{Some Known Statistics}

In this section we will show that the formalism previously
developed permit us  to consider, in a unitary way, the already
known statistical distributions. For simplicity we will discuss
the case of distributions depending exclusively on the velocity.
Firstly we observe that the stationary distribution $f$, defined
through $\kappa(f)=\exp (-\epsilon)$ with $\epsilon = \beta
(mv^2/2-\mu)$, can be obtained as steady state of a Fokker-Planck
(FP) equation describing the kinetics of brownian particles for
which results $U=mv^2/2$ and $D=const$. The same distribution can
be viewed as steady state of a Boltzmann equation, describing free
particles interacting by means of binary collisions, conserving
the particle number, momentum and energy. In this section we will
write the evolution equations (FP and/or Boltzmann) of some
distributions available in literature to illustrate the relevance
of the approach adopted in the previous sections describing the
non linear particle kinetics.

{\it Maxwell-Boltzmann statistics:} We start by considering the MB
statistics given by $f= Z^{-1}\exp (- \epsilon)$. It is readily
seen that the related kinetics is defined starting from $a(f)=f$,
$b(f')=1$, while the symmetric function $c(f,f')$ remains
arbitrary. Then we have $\kappa(f)=f$ and $\gamma(f)=f c(f)$. In
the Boltzmann picture the evolution equation becomes
\begin{eqnarray}
\frac{\partial f}{\partial t} \! =\! \int_{\cal R}\!\!
d^nv'd^nv_1d^nv'_1 \ T c(f,f')c(f_1,f'_1) \big ( f'f'_1\! -\! f
f_1 \big ),
\end{eqnarray}
while, in the Fokker-Planck picture we have
\begin{equation}
\frac{\partial f}{\partial t} = \frac{\partial}{\partial
\mbox{\boldmath$v$}} \! \left[D c(f)\!\left(\!\beta
m\mbox{\boldmath$v$}f+\frac{\partial f}{\partial
\mbox{\boldmath$v$}}\right) \! \right ] \ \ .
\end{equation}
In the simplest case  $c(f,f')=1$ we obtain the standard linear
Boltzmann and FP equations. We observe that there is an infinity
of ways (one for any choice of $c(f,f')$) to obtain the MB
distribution.

{\it Bosonic and fermionic statistics:} We consider now the case
of quantum statistics namely the Fermi-Dirac ($\eta=-1$) and
Bose-Einstein ($\eta=1$) statistics defined by means of $f=
Z^{-1}(\exp \epsilon -\eta)^{-1}$. The kinetics now is defined
through $a(f)=f$ and $b(f')=1+\eta f'$ while again the function
$c(f,f')$ remains arbitrary. We have consequently
$\kappa(f)=f/(1+\eta f)$. In the Boltzmann picture the evolution
equation becomes
\begin{eqnarray}
&&\frac{\partial f}{\partial t}\, = \int_{\cal R}
d^nv'd^nv_1d^nv'_1 \ T c(f,f')c(f_1,f'_1)  \nonumber
\\ && \times  \big [ f'(1 +\eta f ) f'_1(1 +\eta f_1)
 - f(1 +\eta f') f_1 (1 +\eta f'_1) \big ] \ ,
\end{eqnarray}
and reduces to the well known Uehling-Uhlenbeck equation if we
choose $c(f,f')=1$. This choice for $c(f,f')$, in the frame of the
FP picture, implies $\gamma(f)=f(1+\eta f)$ and we obtain the
following evolution equation \cite{KQBF}:
\begin{equation}
\frac{\partial f}{\partial t} = \frac{\partial}{\partial
\mbox{\boldmath$v$}} \! \left[D\beta m\mbox{\boldmath$v$}f(1+\eta
f)+D\frac{\partial f}{\partial \mbox{\boldmath$v$}} \! \right ] \
\ . \label{N24}
\end{equation}
In order to show that a kinetics, different from the one described
by (\ref{N24}),  also reproducing the bosonic and fermionic
statistics, exists, we consider $c(f,f')=(1+\eta \sqrt{f
f'})^{-1}$ or alternatively $c(f,f')=[1+\eta (f +f')/2]^{-1}$. It
is easy to verify that in both the cases we have $c(f)=(1+\eta
f)^{-1}$ and $\gamma(f)=f$. Now the evolution equation in the FP
picture becomes \cite{FD}:
\begin{equation}
\frac{\partial f}{\partial t} = \frac{\partial}{\partial
\mbox{\boldmath$v$}} \! \left[D\beta m
\mbox{\boldmath$v$}f+D\frac{1}{\eta}\frac{\partial}{\partial
\mbox{\boldmath$v$}}\ln (1+\eta f) \! \right ] \ \ .
\end{equation}

{\it Intermediate statistics:} The quantum statistics
interpolating between the bosonic and fermionic statistics  has
captured the attention of many researchers in the last few years.
A first example of intermediate statistics can be realized by
considering in the distribution $f= Z^{-1}(\exp \epsilon
-\eta)^{-1}$, previously examined, the parameter $\eta$ as being
continuous: $0\leq \eta \leq 1$. For $\eta\not=\pm1$ we have a
quantum statistics different from the Bose or Fermi statistics. A
second intermediate statistics is the boson-like $(+)$ or
fermion-like $(-)$ quon statistics \cite{PLA}, which can be
obtained easily by posing $a(f)=[f]_{_{\,\scriptstyle q}}$ and
$b(f')=[1\pm f']_{_{\,\scriptstyle q}}$, where
$[x]_{_{\,\scriptstyle q}}=(q^{x}-q^{-x})/2\ln q$ and $q\in {\bf
R}$. If we choose for simplicity $c(f)=c_{q}=2\ln q / (q-q^{-1})$,
the evolution equation in the FP picture becomes \cite{PLA}:
\begin{equation}
\frac{\partial f}{\partial t} = \frac{\partial}{\partial
\mbox{\boldmath$v$}} \! \left(c_{q}\,D\beta
m\mbox{\boldmath$v$}\,[f]_{_{\,\scriptstyle q}}[1\pm
f]_{_{\,\scriptstyle q}}+D \frac{\partial f}{\partial
\mbox{\boldmath$v$}} \, \right ) \ \ .
\end{equation}
A third intermediate statistics is the Haldane-Wu exclusion
statistics which can be obtained starting from the kinetics
defined by setting $a(f)=f$ and $b(f')=(1-g
f'\,)^{g}\,[1+(1-g)f'\,]^{\,1-g}$ with $0\leq g\leq 1$ \cite{NPB}.

{\it Tsallis statistics:} We consider the non extensive
termostatistics introduced by Tsallis \cite{T}. The relevant
distribution $f=Z^{-1}[1-(1-q)\epsilon]^{1/(1-q)}$ can be obtained
naturally starting from the kinetics defined through $\ln
\kappa(f)=(f^{1-q}-1)/(1-q)\equiv \ln_q f$. The Boltzmann equation
(\ref{N1801}) becomes now
\begin{eqnarray}
\frac{\partial f}{\partial t} =&& \! \! \int_{\cal R}\! \!
d^nv'd^nv_1d^nv'_1 \,T \, c(f,f')c(f_1,f'_1)b(f)b(f_1)b(f')b(f'_1)
\nonumber \\ && \times \bigg [\!\exp\, ( \ln_q f'+\ln_q f'_1 )
-\exp\,(\ln_q f+\ln_q f_1 ) \bigg ] \ \ .
\end{eqnarray}
In the FP picture the evolution equation is given by
\begin{equation}
\frac{\partial f}{\partial t} = \frac{\partial}{\partial
\mbox{\boldmath$v$}} \! \left[D \gamma(f)\left(\beta m
\mbox{\boldmath$v$} + f^{-q}\frac{\partial f }{\partial
\mbox{\boldmath$v$}}\,\right) \right ] \ \ ,
\end{equation}
which for $\gamma(f)=f$ reduces to the one proposed in ref.
\cite{PP}.

\section{The \boldmath$\kappa$-deformed Analysis}

In the previous section we have considered some statistical
distributions, quantum or classical, already known in the
literature, depending on one continuous parameter. We will now
turn our attention to the distribution $f=Z^{-1}(\exp \epsilon
-\eta)^{-1}$. We note that this quantum distribution can be viewed
as a deformation of the MB one, which can be recovered as the
deformation parameter $\eta$ approaches to zero. Another classical
distribution which can be obtained by deforming the MB one, is the
Tsallis distribution $f=Z^{-1}[1-(1-q)\epsilon]^{1/(1-q)}$. The MB
distribution emerges again as the deformation parameter
$q\rightarrow1$.

In the present section we will study the main mathematical
properties of a new, one parameter, deformed exponential function,
while in the next section we will consider the induced deformed
statistics. The deformed exponential is indicated by
$\exp_{_{\{{\scriptstyle \kappa}\}}}(x)$, where $\kappa$ denote
the deformation parameter, and we postulate that it obeys the
following condition:
\begin{equation}
\exp_{_{\{{\scriptstyle \kappa}\}}}(x) \exp_{_{\{{\scriptstyle
\kappa}\}}}(-x)=1 \ \ . \label{N25}
\end{equation}

We start by observing that any function $A(x)$ can be written in
the form  $A(x)=A_e(x)+A_o(x)$ where $A_e(x)=A_e(-x)$ is an even
function and $A_o(x)=-A_o(-x)$ an odd one. The condition
$A(x)A(-x)=1$ allows us to express $A_e(x)$ in terms of $A_o(x)$
by means of $A_e(x)=\sqrt{1+A_o(x)^2}$ and consequently write the
function $A(x)$ in the form: $A(x)= \sqrt{1+A_o(x)^2}+A_o(x)$. At
this point it is obvious that a deformed exponential
$\exp_{_{\{{\scriptstyle \kappa}\}}}(x)$ obeying (\ref{N25}) and
depending only on one deformation parameter $\kappa$, so that
$\exp_{_{\{{\scriptstyle \kappa}\}}}(x){\atop\stackrel
{\textstyle\sim} {\scriptstyle \kappa\rightarrow 0}}\exp x$, can
be written as
\begin{equation}
\exp_{_{\{{\scriptstyle \kappa}\}}}(x)=
\left[\sqrt{1+g_{\kappa}(x)^2}+g_{\kappa}(x)\right]^{1/\kappa}\ \
, \label{N26}
\end{equation}
or alternatively as
\begin{equation}
\exp_{_{\{{\scriptstyle \kappa}\}}}(x)= \exp \left(
\frac{1}{\kappa} \,{\rm arcsinh}\, g_{\kappa}(x) \right)\ \ .
\end{equation}
In (\ref{N26}) the generator $g_{\kappa}(x)$ of the deformed
exponential is an arbitrary function depending on the parameter
$\kappa$ and obeying the conditions:
\begin{equation}
g_{\kappa}(-x)=-g_{\kappa}(x) \ \ ; \ \
g_{\kappa}(x){\atop\stackrel{\textstyle\sim}{\scriptstyle
\kappa\rightarrow 0}}\kappa x \ \ .
\end{equation}
Since it will be useful later on, we introduce the inverse
function of $\exp_{_{\{{\scriptstyle \kappa}\}}}(x)$ indicated by
$\ln_{_{\{{\scriptstyle \kappa}\}}}(x)$ and defined through
$\exp_{_{\{{\scriptstyle \kappa}\}}}\!\left[\ln_{_{\{{\scriptstyle
\kappa}\}}}(x)\right]=x$. It is easy to verify that
\begin{equation}
\ln_{_{\{{\scriptstyle \kappa}\}}}(x)= g_{\kappa}^{-1}\left(
\frac{x^{\kappa}-x^{-\kappa}}{2} \right) \ \ , \label{N27}
\end{equation}
where $g_{\kappa}^{-1}(x)$ is the inverse function of
$g_{\kappa}(x)$. We remark that the deformed exponential can be
defined by fixing the expression of the generator $g_{\kappa}(x)$.
We note that by choosing $g_{\kappa}(x)=\sinh\kappa x$ we can
generate the standard undeformed exponential
$\exp_{_{\{{\scriptstyle \kappa}\}}}(x)= \exp x$ and logarithm
$\ln_{_{\{{\scriptstyle \kappa}\}}}(x)= \ln x $ functions.

{\it The $\kappa$-exponential:} In the following we consider the
simplest deformed exponential (in following called
$\kappa$-exponential), which is generated from
$g_{\kappa}(x)=\kappa x$ and is given by
\begin{equation}
\exp_{_{\{{\scriptstyle \kappa}\}}}(x)=
\left(\sqrt{1+\kappa^2x^2}+\kappa x\right)^{1/\kappa}\ \ ,
\end{equation}
or equivalently by
\begin{equation}
\exp_{_{\{{\scriptstyle \kappa}\}}}(x)= \exp \left(
\frac{1}{\kappa} \,{\rm arcsinh}\, \kappa x \right)\ \ .
\end{equation}
Obviously, we have \hbox{$\exp_{_{\{{0}\}}}(x)=\exp x$} and for
\hbox{$\forall x\in{\bf R}$} it results
\hbox{$\exp_{_{\{{\scriptstyle \kappa}\}}}(x)\in {\bf R}^+$}.
Furthermore we have \hbox{$\exp_{_{\{{\scriptstyle
\kappa}\}}}(0)=1$} and \hbox{$\exp_{_{\{{\scriptstyle
-\kappa}\}}}(x)=\exp_{_{\{{\scriptstyle \kappa}\}}}(x)$}, so that
we can consider \hbox{simply} that $\kappa\in {\bf R}^+$. A
relevant property of $\exp_{_{\{{\scriptstyle \kappa}\}}}(x)$ is
that for $\forall a \in {\bf R}$
\begin{equation}
\exp_{_{\{{\scriptstyle \kappa}\}}}(ax)=[\exp_{_{\{{\scriptstyle
a\kappa}\}}}(x)]^{\,a} \ \ .
\end{equation}
Concerning the asymptotic behaviour of $\kappa$-exponential we
easily obtain that \hbox{$\exp_{_{\{{\scriptstyle
\kappa}\}}}(x){\atop\stackrel{\textstyle\sim}{\scriptstyle
x\rightarrow +\infty}}|2\kappa x|^{1/|\kappa|}$} and
\hbox{$\exp_{_{\{{\scriptstyle
\kappa}\}}}(x){\atop\stackrel{\textstyle\sim}{\scriptstyle
x\rightarrow -\infty}}|2\kappa x|^{-1/|\kappa|}$}. We observe that
the deformed exponential is an increasing function \hbox{$d
\exp_{_{\{{\scriptstyle \kappa}\}}}(x)/dx>0$}, $\forall \kappa \in
{\bf R}$. Finally, its concavity  is $d^2 \exp_{_{\{{\scriptstyle
\kappa}\}}}(x)/dx^2>0$ for $|\kappa|\leq 1$ while for $|\kappa|>
1$ we obtain $d^2 \exp_{_{\{{\scriptstyle \kappa}\}}}(x)/dx^2>0$
for $x<x_c$ and $d^2 \exp_{_{\{{\scriptstyle \kappa}\}}}
(x)/dx^2<0$ for $x>x_c$ being $x_c=(\kappa^4-\kappa^2)^{-1/2}$.

{\it The $\kappa$-sum:} Of course if we take into account that
\hbox{${\rm arcsinh}\,x+ {\rm arcsinh}\,y$} \hbox{$ = {\rm
arcsinh}\,(x\sqrt{1+y^2} +y\sqrt{1+x^2})$} we arrive immediately
at the following relationship:
\begin{equation}
\exp_{_{\{{\scriptstyle \kappa}\}}}(x) \exp_{_{\{{\scriptstyle
\kappa}\}}}(y) =\exp_{_{\{{\scriptstyle \kappa}\}}}( x
\oplus\!\!\!\!\!^{^{\scriptstyle \kappa}}\,\,y )\ \ ,
\end{equation}
where the $\kappa$-deformed sum or simply $\kappa$-sum $x
\oplus\!\!\!\!\!^{^{\scriptstyle \kappa}}\,\,y$, is defined
through
\begin{equation}
x \oplus\!\!\!\!\!^{^{\scriptstyle
\kappa}}\,\,y=x\sqrt{1+\kappa^2y^2}+y\sqrt{1+\kappa^2x^2} \ \ .
\end{equation}
We note that the $\kappa$-sum obeys the associative law $(x
\oplus\!\!\!\!\!^{^{\scriptstyle \kappa}}\,\,y)
\oplus\!\!\!\!\!^{^{\scriptstyle \kappa}}\,\,z=x
\oplus\!\!\!\!\!^{^{\scriptstyle \kappa}}\,\,(y
\oplus\!\!\!\!\!^{^{\scriptstyle \kappa}}\,\,z)$, admits a neutral
element \hbox{$x \oplus\!\!\!\!\!^{^{\scriptstyle \kappa}}\,\,0= 0
\oplus\!\!\!\!\!^{^{\scriptstyle \kappa}}\,\,x= x$} and for any
$x$ exists its opposite \hbox{$x \oplus\!\!\!\!\!^{^{\scriptstyle
\kappa}}\,\,(-x)= 0$}. Moreover the commutativity property
\hbox{$x \oplus\!\!\!\!\!^{^{\scriptstyle \kappa}}\,\,y= y
\oplus\!\!\!\!\!^{^{\scriptstyle \kappa}}\,\,x$} holds, so that
the real numbers constitute an abelian group with respect the
$\kappa$-sum. Since it will be useful later on, we define the
$\kappa$-difference as: $x \ominus\!\!\!\!\!^{^{\scriptstyle
\kappa}}\,\,y= x\oplus\!\!\!\!\!^{^{\scriptstyle
\kappa}}\,\,(-y)$.

{\it The $\kappa$-trigonometry:} Firstly we define the
$\kappa$-deformed hyperbolic sine and cosine, starting from
$\exp_{_{\{{\scriptstyle \kappa}\}}}(\pm
x)=\cosh_{_{\{{\scriptstyle \kappa}\}}}(x)\pm
\sinh_{_{\{{\scriptstyle \kappa}\}}}(x)$ which is the
$\kappa$-Euler formula and obtain
\begin{eqnarray}
\sinh_{_{\{{\scriptstyle \kappa}\}}}(x)
=\frac{1}{2}\left[\exp_{_{\{{\scriptstyle \kappa}\}}}(x)
-\exp_{_{\{{\scriptstyle \kappa}\}}}(-x)\right] \ \ , \nonumber \\
\cosh_{_{\{{\scriptstyle \kappa}\}}}(x)
=\frac{1}{2}\left[\exp_{_{\{{\scriptstyle \kappa}\}}}(x)
+\exp_{_{\{{\scriptstyle \kappa}\}}}(-x)\right] \nonumber \ \ .
\end{eqnarray}
It is straightforward to introduce the $\kappa$-hyperbolic
trigonometry which reduces to the ordinary hyperbolic trigonometry
as $\kappa \rightarrow 0$. For instance, the formulas:
\begin{eqnarray}
&&\tanh_{_{\{{\scriptstyle
\kappa}\}}}(x)=\frac{\sinh_{_{\{{\scriptstyle
\kappa}\}}}(x)}{\cosh_{_{\{{\scriptstyle \kappa}\}}}(x)} \ ,
\nonumber \\ &&\cosh_{_{\{{\scriptstyle \kappa}\}}}^2(x)-
\sinh_{_{\{{\scriptstyle \kappa}\}}}^2(x)=1 \ , \nonumber \\
&&\sinh_{_{\{{\scriptstyle \kappa}\}}}(x
\oplus\!\!\!\!\!^{^{\scriptstyle \kappa}}\,\,y
)\!+\!\sinh_{_{\{{\scriptstyle \kappa}\}}}( x
\ominus\!\!\!\!\!^{^{\scriptstyle \kappa}}\,\,y )
\!=\!2\sinh_{_{\{{\scriptstyle \kappa}\}}}(x)
\cosh_{_{\{{\scriptstyle \kappa}\}}}(x), \nonumber
\\ &&\tanh_{_{\{{\scriptstyle \kappa}\}}}(x)+\tanh_{_{\{{\scriptstyle
\kappa}\}}}(y)=\frac{\sinh_{_{\{{\scriptstyle \kappa}\}}}(x
\oplus\!\!\!\!\!^{^{\scriptstyle \kappa}}\,\,y )}
{\cosh_{_{\{{\scriptstyle \kappa}\}}}(x)\cosh_{_{\{{\scriptstyle
\kappa}\}}}(y)} \ , \nonumber
\end{eqnarray}
and so on, still hold true. All the formulas of the ordinary
hyperbolic trigonometry still hold true after expediently
deformed. The deformation of a given trigonometric formula can be
obtained starting from the corresponding undeformed formula,
making in the argument of the trigonometric functions the
substitution $x+y\rightarrow x \oplus\!\!\!\!\!^{^{\scriptstyle
\kappa}}\,\,y$ and obviously $nx\rightarrow x
\oplus\!\!\!\!\!^{^{\scriptstyle \kappa}}\,\,x ...
\oplus\!\!\!\!\!^{^{\scriptstyle \kappa}}\,\,x$ (n times).

\noindent The $\kappa$-De Moivre  formula involving trigonometric
functions having arguments of the type  $r x$ with $r\in{\bf R}$,
assumes the form
\begin{eqnarray}
[\cosh_{_{\{{\scriptstyle \kappa}\}}}\!(x)\!\pm\!
\sinh_{_{\{{\scriptstyle \kappa}\}}}\!(x)
]^{r}\!\!=\!\cosh_{_{\{{\scriptstyle \kappa}/{\scriptstyle
r}\}}}\!(r x)\!\pm\!\sinh_{_{\{{\scriptstyle \kappa}/{\scriptstyle
r}\}}}\!(r x). \nonumber
\end{eqnarray}

{\it The $\kappa$-derivative:} It is important to emphasize that
we can naturally introduce a ${\kappa}$-calculus which reduces to
the usual one as the deformation parameter ${\kappa}\rightarrow
0$. In fact we can define the ${\kappa}$-derivative of course
through
\begin{equation}
D_{_{\{{\scriptstyle \kappa}\}}}f(x)=\frac{d \, f(x)}{d \,
x_{_{\{{\scriptstyle \kappa}\}}}}=\lim_{y\rightarrow x}\frac{f
(x)-f(y)}{\displaystyle{ x \ominus \!\!\!\!\!^{^{\scriptstyle
\kappa}}\,\,y}}\ \ ,
\end{equation}
where the $\kappa$-differential $d \, x_{_{\{{\scriptstyle
\kappa}\}}}\!= \lim_{y\rightarrow x} \displaystyle{ x \ominus
\!\!\!\!\!^{^{\scriptstyle \kappa}}\,\,y}$, after taking into
account the identity \hbox{$(x \oplus\!\!\!\!\!^{^{\scriptstyle
\kappa}}\,\,y)( x\ominus\!\!\!\!\!^{^{\scriptstyle
\kappa}}\,\,y)=x^2-y^2$},  is given by
\begin{eqnarray}
d \, x_{_{\{{\scriptstyle \kappa}\}}}=\lim_{y\rightarrow x} x
\ominus \!\!\!\!\!^{^{\scriptstyle \kappa}}\,\,y=
\lim_{y\rightarrow x} \frac{x^2-y^2}{\displaystyle{x \oplus
\!\!\!\!\!^{^{\scriptstyle \kappa}}\,\,y}}= \frac{d
x}{\displaystyle{\sqrt{1+\kappa^2x^2} }} \nonumber \ \ .
\end{eqnarray}
The function $x_{_{\{{\scriptstyle \kappa}\}}}$ can be easily
calculated after integration
\begin{eqnarray}
x_{_{\{{\scriptstyle \kappa}\}}}=\frac{1}{\kappa} \ln \left(
\sqrt{1+\kappa^2x^2}+\kappa x \right)=\frac{1}{\kappa}\,{\rm
arcsinh}\, \kappa x \ \ ,
\end{eqnarray}
and satisfies the relationship
$x_{_{\{{\scriptstyle
\kappa}\}}}+y_{_{\{{\scriptstyle \kappa}\}}}=(x
\oplus\!\!\!\!\!^{^{\scriptstyle \kappa}}\,\,y
)_{_{\{{\scriptstyle \kappa}\}}}$. We observe that it is possible
to write the ${\kappa}$-derivative in the form
\begin{equation}
\frac{d \, f(x)}{d \, x_{_{\{{\scriptstyle
\kappa}\}}}}=\sqrt{1+\kappa^2x^2}\ \frac{d\,f(x)}{d x} \ \ ,
\end{equation}
from which it appears clearly that the ${\kappa}$-calculus is
governed by the same rules of the ordinary one. For instance, we
can write
\begin{eqnarray}
\frac{d \,\exp_{_{\{{\scriptstyle \kappa}\}}}(x)
}{d\,x_{_{\{{\scriptstyle \kappa}\}}}}= \exp_{_{\{{\scriptstyle
\kappa}\}}}(x) \ \ ; \ \ \frac{d \,\sinh_{_{\{{\scriptstyle
\kappa}\}}}(x) }{d\,x_{_{\{{\scriptstyle \kappa}\}}}}=
\cosh_{_{\{{\scriptstyle \kappa}\}}}(x) \nonumber ,
\end{eqnarray}
and so on. We observe now that the introduction of the function
$x_{_{\{{\scriptstyle \kappa}\}}}$  permit us to write
$\exp_{_{\{{\scriptstyle \kappa}\}}}(x)= \exp
\left(x_{_{\{{\scriptstyle \kappa}\}}} \right)$ and analogously
for the ${\kappa}$-deformed trigonometric functions. For instance,
we have $\sinh_{_{\{{\scriptstyle \kappa}\}}}(x)= \sinh
\left(x_{_{\{{\scriptstyle \kappa}\}}} \right)$,
$\cosh_{_{\{{\scriptstyle \kappa}\}}}(x)= \cosh
\left(x_{_{\{{\scriptstyle \kappa}\}}} \right)$ etc. This property
of the  ${\kappa}$-deformed exponential and trigonometric
functions permits us to consider their Taylor expansion in terms
of the function $x_{_{\{{\scriptstyle \kappa}\}}}$. For instance,
it holds
\begin{eqnarray}
\exp_{_{\{{\scriptstyle
\kappa}\}}}(x)=\sum_{m=0}^{\infty}\frac{1}{m!} \left(
x_{_{\{{\scriptstyle \kappa}\}}} \right)^m  \ \ .
\end{eqnarray}

{\it The $\kappa$-integral:} Starting from the definition of
$\kappa$-derivative, we define the $\kappa$-integral through
\begin{equation}
\int f(x)\,d \, x_{_{\{{\scriptstyle \kappa}\}}}= \int \frac{f(x)}
{\displaystyle{\sqrt{1+\kappa^2x^2}}}\, d x \ \ .
\end{equation}
All the standard rules of the undeformed integral calculus still
hold true.

{\it The $\kappa$-cyclic functions:} The $\kappa$-deformed  sine
and cosine can be defined through $\sin_{_{\{{\scriptstyle
\kappa}\}}}(x)=-i\sinh_{_{\{{\scriptstyle \kappa}\}}}(ix)$ and
$\cos_{_{\{{\scriptstyle \kappa}\}}}(x)=\cosh_{_{\{{\scriptstyle
\kappa}\}}}(ix)$, respectively. It results:
$\sin_{_{\{{\scriptstyle \kappa}\}}}(x)=\sin (x_{_{\{{\scriptstyle
i\kappa}\}}}) $ and $\cos_{_{\{{\scriptstyle
\kappa}\}}}(x)=\cos(x_{_{\{{\scriptstyle i\kappa}\}}}) $ being
$x_{_{\{{\scriptstyle i\kappa}\}}}=\kappa^{-1}\arcsin(\kappa x)$.
Of course the $\kappa$-cyclic trigonometry, naturally can be
introduced.

{\it The $\kappa$-logarithm:} We study now the inverse function of
the ${\kappa}$-exponential namely the ${\kappa}$-logarithm which,
after remembering its definition (\ref{N27}) and the expression of
the generator $g_{\kappa}(x)=\kappa x$, can be written as
\begin{equation}
\ln_{_{\{{\scriptstyle \kappa}\}}}(x)=
\frac{x^{\kappa}-x^{-\kappa}}{2\kappa} \ \ . \label{N28}
\end{equation}
From (\ref{N28}) we can see that $\forall x\in{\bf R}^+\Rightarrow
\ln_{_{\{{\scriptstyle \kappa}\}}}(x) \in {\bf R}$ and
$\ln_{_{\{{0}\}}}(x)=\ln x$. Furthermore we have
$\ln_{_{\{{\scriptstyle \kappa}\}}}(1)=0$ and
$\ln_{_{\{{\scriptstyle -\kappa}\}}}(x)=\ln_{_{\{{\scriptstyle
\kappa}\}}}(x)$. Two relevant properties of the
${\kappa}$-logarithm are $\ln_{_{\{{\scriptstyle
\kappa}\}}}(xy)=\ln_{_{\{{\scriptstyle \kappa}\}}}(x)
\oplus\!\!\!\!\!^{^{\scriptstyle \kappa}}\,\,
\ln_{_{\{{\scriptstyle \kappa}\}}}(y)$ and $\ln_{_{\{{\scriptstyle
\kappa}\}}}(x^m)=m\ln_{_{\{{\scriptstyle m\kappa}\}}}(x)$, while
its asymptotic behaviour is \hbox{$\ln_{_{\{{\scriptstyle
\kappa}\}}}(x){\atop\stackrel{\textstyle\sim}{\scriptstyle
x\rightarrow {\,0^+}}}-|2\kappa|^{-1}x^{-|\kappa|}\rightarrow
-\infty$} and \hbox{$\ln_{_{\{{\scriptstyle
\kappa}\}}}(x){\atop\stackrel{\textstyle\sim}{\scriptstyle
x\rightarrow +\infty}}|2\kappa|^{-1}x^{|\kappa|}\rightarrow
+\infty$}.

In closing this section we consider the relationship linking the
functions $\exp_{_{\{{\scriptstyle \kappa}\}}}(x)$ and
$\ln_{_{\{{\scriptstyle \kappa}\}}}(x)$, here introduced with the
already known in the literature Tsallis exponential
$\exp_{q}(x)=[1+(1-q)x]^{1/(1-q)}$ and its inverse function, the
Tsallis logarithm $\ln_{q}(x)=(x^{1-q}-1)/(1-q)$ namely:
\begin{eqnarray}
&&\exp_{_{\{{\scriptstyle \kappa}\}}}(x)= \exp_{_{\scriptstyle
1+\kappa}}\left(x+\frac{\sqrt{1+\kappa^2x^2}-1}{\kappa}\right) \ \
, \nonumber \\ &&\ln_{_{\{{\scriptstyle
\kappa}\}}}(x)=\frac{1}{2}\left[ \ \ln_{_{\scriptstyle
1+\kappa}}(x)+\ln_{_{\scriptstyle 1-\kappa}}(x)\,\right] \ \ .
\nonumber
\end{eqnarray}

\section{The  \boldmath$\kappa$-deformed Statistics}

In this section we will consider a new statistical distribution,
just as a working example of the theory presented in the previous
sections. We start by considering a particle system in the
velocity space and postulating the following density of entropy:
\begin{equation}
\sigma_{_{\scriptstyle \kappa}}(f)= - \int d f \
\ln_{_{\{{\scriptstyle \kappa}\}}}(\alpha f) \ \ , \label{N29}
\end{equation}
where the real constant $\alpha>0$ is not specified for the
moment. Eq. (\ref{N29}), for $\kappa\rightarrow 0$, gives the
standard Boltzmann-Gibbs-Shannon density of entropy if we pose
$\alpha =1$. We note that the above definition of
$\sigma_{_{\scriptstyle \kappa}}(f)$ implies that $\kappa (f)$ is
related to the $\kappa$-logarithm, through $\ln
\kappa(f)=\ln_{_{\{{\scriptstyle \kappa}\}}}(\alpha f)$. It is
immediate to see that for $\forall \kappa \in{\bf R}$, $d^2
\,\sigma_{_{\scriptstyle \kappa}}(f)/d \, f^2\leq 0$,
independently on the value of $\alpha$. Then for the entropic
functional ${\cal K}$, we have $d{\cal K}(t)/dt\geq 0$ and ${\cal
K}(t)\leq{\cal K}(\infty)$. The entropy of the system given by
$S_{_{\scriptstyle \kappa}}\!\!=\!\!\int_{\cal R}
d^nv\,\sigma_{_{\scriptstyle \kappa}}(f)$ assumes now the form
\begin{equation}
S_{_{\scriptstyle \kappa}}=-\frac{1}{2\kappa}\,\int_{\cal R} d^nv
\,\left(\, \frac{\alpha^{\,\kappa}}{1\!+\!\kappa}\,
f^{\,1+\kappa}- \frac{\alpha^{\!-\kappa}}{1\!-\!\kappa}\,
f^{\,1-\kappa} \right) \ , \label{N30}
\end{equation}
and reduces to the standard Boltzmann-Gibbs-Shannon  entropy $
S_{_{\scriptstyle 0}}=-\int_{\cal R} d^n v \,[\,\ln (\alpha f)
 \!-\!1]f$ as the deformation parameter approaches zero.
This $\kappa$-entropy is linked to the Tsallis entropy
$S^{^{(T)}}_{q}$ through the following relationship

\begin{equation}
S_{_{\scriptstyle \kappa}}=\frac{1}{2}\,
\frac{\alpha^{\,\kappa}}{1\!+\!\kappa}\,
S^{^{(T)}}_{_{\scriptstyle 1+\kappa}} +
\frac{1}{2}\,\frac{\alpha^{\!-\kappa}}{1\!-\!\kappa}\,
S^{^{(T)}}_{_{\scriptstyle 1-\kappa}} + const. \label{N300}
\end{equation}

{\it First choice of $\alpha$}: We discuss now the case
$\alpha=1$. The stationary statistical distribution corresponding
to the entropy $S_{_{\scriptstyle \kappa}}$ can be obtained by
maximizing the functional ${\cal K}$
\begin{equation}
\delta\left[ S_{_{\scriptstyle \kappa}}+\int_{\cal R} d^n
v(\beta\mu f-\beta U\! f)\right] =0 \ \ .
\end{equation}
One arrives to the following distribution
\begin{equation}
f=\exp_{_{\{{\scriptstyle \kappa}\}}}(-\epsilon) \ \ ; \ \
\epsilon=\beta(U-\mu) \ \ , \label{N34}
\end{equation}
which gives the standard classical statistical distribution as
$\kappa\rightarrow 0$.

{\it Second choice of $\alpha$}: Before introducing the second
choice of the constant $\alpha$ we consider briefly the concept of
entropy. It is to be understood that the entropy is an absolute
measure of information only in the case of an isolated system,
where the particle number and energy are conserved. In the case of
a system interacting with a bath, only the particle number is
conserved and the free entropy can't be used as an absolute
measure of information. For this reason an entropy constrained by
the particle number conservation and by the relevant energy mean
value must be introduced. The constrain introduced by the particle
number conservation is a special one, and because of its presence
both in the case of free and interacting systems, can be taken
into account directly in the definition of the entropy. For
instance, for the linear kinetics, the stationary distribution
$f=Z^{-1}\exp(-\beta U)$ with partition function given by
$Z\!=\!\int_{\cal R} d^nv \, \exp(-\beta U)$, can be obtained
using the variational principle, namely $\delta [
S_{_{\scriptstyle 0}}-\beta\int_{\cal R} d^n v\, U f ] =0$, where
the functional $ S_{_{\scriptstyle 0}}\!=\! -\!\int_{\cal R} d^n v
\,[\,\ln (Z f)-1]\,f$ depending on the constant $Z$ is the above
mentioned constrained entropy, the particle number (in following
we pose $N=1$) being conserved.

It is clear that, for analogy, also in the case of the non linear
kinetics with $\kappa\not=0$, we can choose $\alpha=Z$ in the
expression of $S_{_{\scriptstyle \kappa}}$ given by (\ref{N30}),
so that the stationary statistical distribution
\begin{equation}
f=\frac{1}{Z}\exp_{_{\{{\scriptstyle \kappa}\}}}(-\beta U) \ \ ,
\label{N341}
\end{equation}
with $Z=\int_{\cal R} d^nv\, \exp_{_{\{{\scriptstyle
\kappa}\}}}(-\beta U)$  can be obtained by considering the
following variational principle
\begin{equation}
\delta\left[ S_{_{\scriptstyle \kappa}}-\beta\int_{\cal R} d^n v\,
U\, f\,\right] =0 \ \ .
\end{equation}

Of course, the expression of $f$ depends on the potential $U$ and,
in the particularly interesting case of Brownian particles, $U=m
v^2/2$, after tedious but straightforward calculations we can
write the distribution function as:

\begin{equation}
\!\!\!\!\!\!f\!=\!\!\left[\!\frac{\beta m
|\kappa|}{\pi}\right]^{\!\!\textstyle{ \frac{n}{2}}} \!\! \left
[\!1\!\!+\!\!\frac{1}{2}n|\kappa| \right ]\!
\!\frac{\Gamma\!\left(\!\textstyle{\frac{1}{2|\kappa|}\!+
\!\frac{n}{4}}\!\right)}{\Gamma\!\left(\!\textstyle{\frac{1}{2|\kappa|}\!
 -\!\frac{n}{4}}\! \right)}\exp_{_{\{{\scriptstyle
\kappa}\}}}\!\!\left(\!\! -\frac{\beta}{2}m v^2\! \!\,\right)\!
,\! \! \label{N31}
\end{equation}
where \hbox{$n=1,2,3$} is the dimension of the velocity space and
\hbox{$|\kappa|<2/n$}. The distribution given by (\ref{N31})
reduces to the standard Maxwell-Boltzmann one \hbox{$f=(\beta
m/2\pi)^{n/2}\exp (-\beta m v^2/2)$} as the deformation parameter
$\kappa$ approaches to zero. This limit can be performed easily by
taking into account the Stirling's formula:
\hbox{$\,\Gamma(z)\!\sim\! \sqrt{2\pi}\,z^{z-1/2}\exp(-z)$},
holding for \hbox{$z\!\rightarrow\!+\infty$}.

We write now the evolution equation, whose stationary state is
described by (\ref{N31}). After indicating with $f(t,
\mbox{\boldmath$v$})$ the time depending statistical distribution,
which for $t\!\rightarrow\!\infty$ reduces to  (\ref{N31}), we
introduce the new function $p=Z_{\infty}f$, and remember that
$\alpha=Z_{\infty}$. The evolution equation of the function $p$,
in the Fokker-Planck picture, in the case of Brownian particles,
and by choosing for simplicity $\gamma (f)=f$, assumes the form
\begin{equation}
\frac{\partial p}{\partial t} = \frac{\partial}{\partial
\mbox{\boldmath$v$}} \! \left[k \mbox{\boldmath$v$}
\,p+\frac{D}{2}\, (\,p^{\kappa}+p^{-\kappa})\, \frac{\partial
p}{\partial \mbox{\boldmath$v$}} \right ],
\end{equation}
where $k=Dm\beta_{\infty}$. In the Boltzmann picture the evolution
equation is structurally similar with the one of Tsallis kinetics.
The only difference is that the Tsallis logarithm is replaced now
by the $\kappa$-logarithm.

We conclude this section by considering a new quantum
distribution, describing particles with a behaviour intermediate
between bosons and fermions, which can be constructed starting
from (\ref{N34}). We impose for the entropy density the following
expression:
\begin{equation}
\sigma_{_{\scriptstyle \kappa}}(f)= - \int  d f \
\ln_{_{\{{\scriptstyle \kappa}\}}}\left(\frac{f}{1+\eta f}\right)
\ \ , \label{N32}
\end{equation}
being $\eta$ a real parameter. After maximization of the
constrained entropy or, equivalently, after obtaining the
stationary solution of the proper evolution equation associated to
(\ref{N32}), one arrives to the following distribution
\begin{equation}
f= \frac{1}{\exp_{_{\{{\scriptstyle \kappa}\}}}(\epsilon)-\eta} \
\ . \label{N33}
\end{equation}
We note that (\ref{N33}) for $\eta=0$ reduces to (\ref{N34}) while
for $\eta=1$ becomes a $\kappa$-deformed Bose-Einstein
distribution and for $\eta=-1$ becomes a $\kappa$-deformed
Fermi-Dirac distribution. Finally for $\eta\neq 0,\pm 1$ the
(\ref{N33}) becomes a quantum distribution which defines an
intermediate statistics and can describe anyons likewise of the
distribution considered in ref. \cite{AN,KQBF}. Obviously, other
quantum intermediate statistical distributions (generalized
Haldane statistics,   generalized quon statistics, etc.), can be
constructed starting from (\ref{N34}).

\section{Applications of $\kappa$-deformed statistics}

It is worth to note here that $\kappa-$deformed  statistical
distribution given by (\ref{N34}) and (\ref{N341}) is obtained by
extremization, under constraints, of the entropy
$S_{_{\scriptstyle \kappa}}$ which is given by (\ref{N300}) in
terms of Tsallis entropy. This interesting result permits us to
use $\kappa-$deformed  statistical distribution to study physical
systems where we can find experimental evidences for the physical
relevance of the Tsallis entropy like, for instance, in 2d
turbolent pure-electron plasma \cite{BOG}, solar neutrinos
\cite{NEU,NEU1,NEU2}, bremsstrahlung \cite{BRE}, anomalous
diffusion of correlated and Levy type \cite{BO,AD1},
self-gravitating systems \cite{SG}, cosmology \cite{GAL} among
many others. In the following we consider two examples of
application of $\kappa$-deformed distribution, just to see the
values the parameter $\kappa$ assumes.

In ref. \cite{NEU,NEU1,NEU2}, firstly,  the problem of the solar
neutrinos is considered and a solution, in the frame of the non
extensive statistics is proposed. It is well known that the solar
core, where the energy is produced, is a weakly non ideal plasma.
In fact, density and temperature condition suggest that the
microscopic diffusion of ions is non standard: diffusion and
friction coefficients are energy dependent, collisions are not
two-body processes and retain memory beyond the single scattering
event. For this reason the equilibrium velocity distribution of
the ions is slightly different in the tail from the Maxwellian
one, as argued also by Clayton \cite{CLA,CLA1}. Consequently the
reaction rates are sensibly modified and, at last, the neutrino
fluxes which are experimentally detected, have calculated values
different from the standard ones. With the hypothesis that the
velocity distribution of the ions in the solar core is a
$\kappa$-deformed distribution, we can reproduce the experimental
data, analyzed in ref. \cite{NEU,NEU1,NEU2}, when $\kappa=0.15$.

In ref. \cite{GAL} it is shown that the observational data
concerning the velocity distribution of clusters of galaxies can
be fitted by a  non extensive statistical distribution. If we
adopt the $\kappa$-deformed statistical distribution to analyze
the data reported in ref. \cite{GAL} we obtain a remarkable good
fitting when $\kappa=0.51$.

\section{Conclusions}

In the present effort we have studied from a general prospective
the kinetics of non linear systems in the frame of two pictures,
namely Kramers and Boltzmann. The results, can be summarized as
follows:

The KIP governs the time evolution of the non linear system by
means of the function $\gamma(f,f')= a(f) \, b(f')\, c(f,f')$ and
imposes its steady state through the function
$\kappa(f)=a(f)/b(f)$. The steady state $f_{\!s}$ can be obtained
as stationary solution of its evolution equation.

The KIP imposes the entropy form of the non linear system, which
is given by (\ref{M5}) both in Kramers and in Boltzmann pictures.
The constrained entropy ${\cal K}$, given by (\ref{M3}), satisfies
the H theorem when $d\kappa(f)/df \geq 0$ and the $f_{\!s}$ can be
obtained also from the maximum entropy principle for ${\cal K}$.

In the case of Brownian particles where
$U(\mbox{\boldmath$v$})=\frac{1}{2}m \mbox{\boldmath$v$}^2$, the
stationary state $f_{\!s}$ and then the constrained entropy ${\cal
K}$ assume the same values both in Kramers and in Boltzmann
pictures. Being $\kappa(f)$ an arbitrary function, the H-theorem
has been demonstrated in a unified way for a very large class of
isolated or interacting with a bath non linear systems.

Finally, we have considered within the formalism here developed,
some statistical distributions already known in the literature. On
the other hand, as a working example of the theory here presented,
we have introduced the new $\kappa$-deformed statistics. After
studying the main properties of the new statistics, we have
discussed two applications to real physical systems.

\end{document}